# ACCELERATORS IN MACROECONOMICS: COMPARISON OF DISCRETE AND CONTINUOUS APPROACHES

**Valentina V. Tarasova,**
Lomonosov Moscow State University Business School,
Lomonosov Moscow State University, Moscow 119991, Russia
E-mail: v.v.tarasova@mail.ru
**Vasily E. Tarasov**
Skobeltsyn Institute of Nuclear Physics,
Lomonosov Moscow State University, Moscow 119991, Russia
E-mail: tarasov@theory.sinp.msu.ru

**Abstract:** We prove that the standard discrete-time accelerator equation cannot be considered as an exact discrete analog of the continuous-time accelerator equation. This leads to fact that the standard discrete-time macroeconomic models cannot be considered as exact discretization of the corresponding continuous-time models. As a result, the equations of the continuous and standard discrete models have different solutions and can predict the different behavior of the economy. In this paper, we propose a self-consistent discrete-time description of the economic accelerators that is based on the exact finite differences. For discrete-time approach, the model equations with exact differences have the same solutions as the corresponding continuous-time models and these discrete and continuous models describe the same behavior of the economy. Using the Harrod-Domar growth model as an example, we show that equations of the continuous-time model and the suggested exact discrete model have the same solutions and these models predict the same behavior of the economy.

## 1. Introduction

Economic accelerator is a fundamental concept of macroeconomic theory (Allen, 1960; Allen, 1968). Accelerators can be considered in the models with continuous and discrete time. The continuous-time accelerators are described by using equations with derivative of the first order. The discrete-time accelerators are described by using the equations with finite differences. One of the simplest macroeconomic models, in which the concept of the accelerator is used, is the Harrod-Domar growth model proposed in works (Harrod, 1936; Domar, 1946; Domar, 1947). The Harrod-Domar growth model with continuous time (Allen, 1960, p. 64-66) and the Harrod-Domar growth model with discrete time (Allen, 1960, p. 74-76) are not equivalent. A similar situation occurs with other macroeconomic models. The discrete-time macroeconomic models cannot be considered as exact discrete analogs of continuous-time models. The equations of these models have different solutions and can predict the different behavior of the economy. In this regard, it is important to understand the reasons for the lack of equivalence of discrete and continuous models.

It is well-known that the standard finite differences of integer orders cannot be considered as an exact discretization of the integer derivatives. Therefore the discrete-time accelerator equation with the standard finite differences cannot be considered as an exact discrete analog of the accelerator equation, which contains the derivative of first order. To define discrete-time accelerators that are exact discrete analogs of continuous-time accelerators, we should consider an exact correspondence between the continuous and discrete time approaches. The problem of exact discretization of the differential equations of integer orders has been formulated by Potts (1982; 1986) and Mickens (1988, 1993, 1999, 2000, 2002, 2005). It has been proved that for differential equations there is a finite-difference discretization such that the local truncation errors are zero. A main disadvantage of this approach to discretization is that the suggested differences depend on the form of the type and parameters of the considered differential equation. In addition, these differences do not have the same algebraic properties as the integer derivatives. Recently, a new approach to the exact discretization has been suggested by Tarasov (2014, 2015a, 2015b, 2016a, 2016b, 2017). This approach is based on the principle of universality and the algebraic correspondence principle (Tarasov, 2016a). The exact finite differences have a property of universality if they do not depend on the form and parameters of the considered differential equations. An algebraic correspondence means that the exact finite differences should satisfy the same algebraic relations as the derivatives. In this paper, we propose a self-consistent discrete-time description of the economic accelerators that is based on the exact finite differences.

## 2. Accelerator

In macroeconomics, the accelerator describes how much the change in the value of the endogenous variable (for example, the induced investment I(t)) in response of a single relative increase of the exogenous variable (for example, the income Y(t)). The formulation of the accelerator depends on whether continuous or discrete analysis is used. The simplest expression of the linear accelerator in the continuous form without lags (Allen, 1960, p. 62-63) has the form

$$I(t) = v \cdot \frac{dY(t)}{dt}, \quad (1)$$

where $dY(t)/dt$ is the rate of output (income), and $I(t)$ is the rate of induced investment, each as a flow at time t, and v is a positive constant, the investment coefficient indicating the power of the accelerator. Equation (1) means that induced investment is here a constant proportion of the current rate of change of output.

In discrete analysis, the linear accelerator without lags can be written (Allen, 1960, p. 63) in the form

$$I_t = v \cdot (Y_t - Y_{t-1}), \quad (2)$$

in which the unit step (T=1) is supposed and $Y_t = Y(t)$ for integer values of t. This discrete equation corresponds to the equation (1). In the discrete approach with an arbitrary step T>0, the linear accelerator can be written in the form

$$I_n = \frac{v}{T} \cdot (Y_n - Y_{n-1}), \quad (3)$$

where $Y_n = Y(n \cdot T)$, $I_n = I(n \cdot T)$, and T is a positive constant indicating the time scale. If T=1, then t=n and $Y_n = Y_t$. In this case, equation (3) takes the form (2). Equations (2) and (3) mean that induced investment depends on the current change in output (Allen, 1960, p. 63).

Using the standard finite differences, such as the backward difference $\Delta_b^1 Y(t) := Y(t) - Y(t-1)$, equation (2) can be written as

$$I(t) = v \cdot \Delta_b^1 Y(t). \tag{4}$$

Equations (2), (3) and (4) cannot be considered as exact discrete analogs of equation (1). This is caused by that the standard finite differences, such as the backward difference $\Delta_b^1 Y(t) := Y(t) - Y(t-1)$ and forward difference $\Delta_f^1 Y(t) := Y(t+1) - Y(t)$, do not have the same basic characteristic properties as the derivatives of first order (Tarasov, 2015a; Tarasov, 2016a). For example, the standard Leibniz rule (the product rule), which is a characteristic property of derivatives, is violated for the standard finite differences (Tarasov, 2015a; Tarasov, 2016a), that is, we have the inequality

$$\Delta_b^1 \left( X_1(t) \cdot X_2(t) \right) \neq \left( \Delta_b^1 X_1(t) \right) \cdot X_2(t) + X_1(t) \cdot \left( \Delta_b^1 X_2(t) \right). \tag{5}$$

For the backward difference, the product rule has the nonstandard form

$$\Delta_b^1 \left( X_1(t) \cdot X_2(t) \right) = \left( \Delta_b^1 X_1(t) \right) \cdot X_2(t) + X_1(t) \cdot \left( \Delta_b^1 X_2(t) \right) - \left( \Delta_b^1 X_1(t) \right) \cdot \left( \Delta_b^1 X_2(t) \right). \tag{6}$$

For comparison, we give the action of the derivative and the standard finite difference on some elementary functions in the form of Table 1.

| f(t) | df(t)/dt | $\Delta_b^1 f(t)$ |
|---|---|---|
| $\exp(\lambda \cdot t)$ | $\lambda \cdot \exp(\lambda \cdot t)$ | $\dfrac{\exp(\lambda) - 1}{\exp(\lambda)} \cdot \exp(\lambda \cdot t)$ |
| $\sin(\lambda \cdot t)$ | $\lambda \cdot \cos(\lambda \cdot t)$ | $2 \cdot \sin\left(\lambda \cdot t - \dfrac{\lambda}{2}\right) \cos\left(\dfrac{\lambda}{2}\right)$ |
| $\cos(\lambda \cdot t)$ | $-\lambda \cdot \sin(\lambda \cdot t)$ | $-2 \cdot \sin\left(\lambda \cdot t - \dfrac{\lambda}{2}\right) \sin\left(\dfrac{\lambda}{2}\right)$ |
| $t^2$ | $2 \cdot t$ | $2 \cdot t - 1$ |
| $t^3$ | $3 \cdot t^2$ | $3 \cdot t^2 - 3 \cdot t + 1$ |

Table 1: Actions of derivatives and standard finite differences.

We can see that the action of standard difference $\Delta_b^1$ does not coincide with the action of first derivative in general. As a result, in the general case the solutions of the equations with standard finite differences do not coincide with solutions of the differential equations, which are derived by the replacement of the standard finite differences by the derivatives of the same orders (Tarasov, 2016a).

The nonequivalence of the action of derivatives and standard finite differences leads to the fact that macroeconomic models with discrete time are not equivalent to the corresponding models with continuous time. In the next section, we demonstrate the nonequivalence of the continuous and discrete macroeconomic models by using the Harrod-Domar growth models.

### 3. Harrod-Domar Growth Models

*3.1 Continuous Time Approach*

Let us consider the Harrod-Domar growth model with continuous time (Allen, 1960, p. 64-66). If autonomous investment A(t) grow, for example, as a result of the sudden appearance of large inventions, the multiplier gives a corresponding increase $A(t)/(1-c)$ in output, where c is the marginal value of propensity to consume (0<c<1). The expansion of output activates the accelerator and leads to further (induced) investment. These additional investments increase output due to the multiplier effect and another cycle begins.

The Harrod-Domar model describes the interaction of the multiplier and the accelerator in the absence of delays (lags) and the simplest form of an accelerator. In a continuous time approach, all variables are taken as continuous functions of time and relations are assumed linear. If we select independent (autonomous) expenditures for both consumption and capital investment, the basic condition (balance equation) can be written in the form

$$Y(t) = C(t) + I(t) + A(t), \qquad (7)$$

where $Y(t)$ is the output (income), $C(t)$ is the consumption, $I(t)$ is the induced investment, and $A(t)$ is the autonomous investment. Here we can use the consumption function $C(t) = c \cdot Y(t)$ and accelerator equation (1) with 0<c<1 and $v > 0$. As a result, we get the equation

$$Y(t) = c \cdot Y(t) + v \cdot \frac{dY(t)}{dt} + A(t). \qquad (8)$$

Equation (8) can be rewritten in the form

$$\frac{dY(t)}{dt} = \lambda \cdot Y(t) - \frac{1}{v} \cdot A(t), \qquad (9)$$

where $\lambda = s/v$ and $s = 1 - c$ is the marginal propensity to save. Equation (9) is the differential equation, whose solution described the dynamics of output $Y(t)$ over time. The solution of (9) depends on the dynamics of autonomous expenditure $A(t)$ over time. Let us consider the case of the fixed autonomous expenditure ($A(t) = A = \text{const}$). Let $y(t)$ be the deviation of income from the fixed level $A/s$, i.e., $y(t) = Y(t) - A/s$ and $dy(t)/dt = dY(t)/dt$. Then equation (9) can be rewritten in the form

$$\frac{dy(t)}{dt} = \lambda \cdot y(t), \qquad (10)$$

where $\lambda = s/v$. The solution of equation (10) has the form

$$y(t) = y(0) \cdot \exp(\lambda \cdot t), . \qquad (11)$$

where $y(0)$ is a constant that described the initial income level. Using $y(t) = Y(t) - A/s$, we get the solution of equation (9) with $A(t) = A$ in the form

$$Y(t) = A/s + (Y(0) - A/s) \cdot \exp(\lambda \cdot t). \qquad (12)$$

Solution (12) expresses continuous growth of output or income with a constant growth rate $\lambda = s/v > 0$. Usually the marginal propensity to save $s = 1 - c$ is quite small in comparison with the investment coefficient v. In this case the growth rate $\lambda = s/v$ is a positive fraction that may be quite small.

*3.2. Discrete Time Approach*

Let us consider the Harrod-Domar growth model with discrete time (Allen, 1960, p. 74-76). The main Harrod assumption is that saving plans, rather than consumption plans, are realized. This is one possible assumption that leads to the introduction of delays. In the linear case, when we exclude any autonomous expenditure $A_t = 0$, the saving function has the form $S_t = s \cdot Y_{t-1}$, where s is the constant marginal propensity to save. Generally speaking, this is the expected ratio. But, as savings plans are implemented, $S_t$ is also the actual value of savings. The expected consumption will be equal to $(1-s) \cdot Y_{t-1}$, the actual consumption is determined by the formula

$$C_t = Y_t - S_t = Y_t - s \cdot Y_{t-1}. \tag{13}$$

The balance equation, which connects the actual values of the model, is analogous to equation (7) of continuous model and it has the form

$$Y_t = C_t + I_t + A_t, \tag{14}$$

where $I_t$ is induced investment, and $A_t$ is independent investment. Therefore, we have equation $S_t = I_t + A_t$, which expresses the actual equality of savings and investment. Let us consider the most important case, when there are no autonomous investments. For $A_t = 0$, the actual investment, which all are induced, is given by expression

$$I_t = S_t = s \cdot Y_{t-1}. \tag{15}$$

Expected induced investments express the action of the accelerator without lag (delay) in the form

$$J_t = v \cdot (Y_t - Y_{t-1}). \tag{16}$$

The further specification of the model depends on the relationship between the expected investment $J_t$ and actual investments $I_t$. The growth rate of output $Y_t$ is given by the equilibrium condition that investment plans are always realized ($J_t = I_t$) for all t. Since saving plans are assumed realized in the first place, this is the special type of situation in which saving and investment are always the same, expected and actual. This condition is expressed by the equation

$$v \cdot (Y_t - Y_{t-1}) = s \cdot Y_{t-1}. \tag{17}$$

Using the backward difference $\Delta_b^1 Y(t) := Y(t) - Y(t-1)$, equation (17) can be written in the form

$$\Delta_b^1 Y(t) = \lambda \cdot Y_{t-1}, \tag{18}$$

where $\lambda = s/v$. Equation (17) also can be written in the form $Y_t = (1 + \lambda) \cdot Y_{t-1}$. The solution of this difference equation (Allen, 1960, p. 76) has the form

$$Y_t = Y_0 \cdot (1 + \lambda)^t = Y_0 \cdot \exp(t \cdot \ln(1 + \lambda)). \tag{19}$$

Solution (19) expresses continuous growth of output or income with the constant relative speed $\ln(1 + \lambda)$.

Let us consider the case of the fixed autonomous expenditure ($A(t) = A = \text{const}$). The equation has the form

$$\Delta_b^1 Y(t) = \lambda \cdot Y_{t-1} - \frac{A}{v}. \tag{20}$$

The solution of equation (20) can be given (Allen, 1960, p. 185-186) by the expression

$$Y_t = A/s + (Y_0 - A/s) \cdot \exp(t \cdot \ln(1 + \lambda)). \tag{21}$$

which described the growth of income with the constant growth rate $\ln(1 + \lambda)$.

If we take into account the step $T \neq 1$, solution (19) takes the form $Y_t = Y_0 \cdot (1 + \lambda \cdot T)^{t/T}$. Only in the limit $T \to 0$, we get $Y(t) = Y(0) \cdot \exp(\lambda \cdot t)$, by using $\lim_{x \to 0} (1 + x)^{1/x} = e$. It is easy to see by direct substitution that the expression (12) is not a solution of the difference equation (20) since $\Delta_b^1 \exp(\lambda \cdot t) \neq \lambda \cdot \exp(\lambda \cdot t)$.

As a result, we can see that the growth rate $\ln(1 + \lambda)$ of the discrete models does not coincide with growth rate $\lambda = s/v$ of the continuous model.

The similar situation occurs with other macroeconomic growth models, including the natural growth model, the Keynes model, the dynamic intersectoral model of Leontief, and others.

Using the Harrod-Domar growth model as an example, we show that the discrete-time macroeconomic models, which are based on standard differences, cannot be considered as exact discrete analogs of continuous-time models. The equations of these models can have different solutions and can predict the different behavior of the economy. In the next section, we propose a self-consistent discrete-time description of the economic accelerators that allows us to propose discrete macroeconomic models, which can be considered as exact discretization of the corresponding continuous-time models. In addition these discrete models predict the same behavior of the economy as the corresponding continuous-time macroeconomic models.

## 4. Concept of Exact Discretization

In order to have difference equations of the accelerator, which can be considered as exact discrete analogs of equation (1), we propose to use the requirement on difference operators in the form of the correspondence principle (Tarasov, 2016a): The finite differences, which are exact discretization of derivatives of integer orders, should satisfy the same algebraic characteristic relations as these derivatives. The suggested principle of algebraic correspondence means that the correspondence between the discrete and continuous time economic models lies not so much in the limiting condition, when the step tends to zero (T → 0) as in the fact that mathematical operations on these two models should obey in many cases the same mathematical laws.

The exact discrete analogs of the derivatives should have the same basic characteristic properties as these derivatives (Tarasov, 2016a):

(1) The Leibniz rule is a characteristic property of the derivatives of integer orders. Therefore the exact discretization of the derivatives should satisfy this rule. The Leibniz rule should be the main characteristic property of exact discrete analogs of the derivatives.

(2) The exact discretization should satisfy the semi-group property. For example, the exact finite difference of second-order should be equal to the repeated action of the exact differences of the first order.

(3) The exact differences of power-law functions should give the same expression as an action of derivatives. This allows us to consider the exact correspondence of derivatives and differences on the space of entire functions.

Tarasov (2014, 2015a, 2015b, 2016a, 2016b, 2017) we proposed new approach to exact discretization that is based on new difference operators, which can be considered as an exact discretization of derivatives of integer and non-integer orders. These differences do not depend on the form and parameters of considered differential equations. Using these differences, we can get an exact discretization of differential equation of integer and non-integer orders. The suggested approach to exact discretization allows us to obtain difference equations that exactly correspond to the differential equations. We consider not only an exact correspondence between the equations, but also exact correspondence between solutions. The suggested exact differences allow us to propose the exact discrete-time analogs of the continuous-time equations of the accelerators.

## 5. Exact Discrete Analogs of Standard Accelerators

Let E(R) be a space of entire function on the real axis R and E(Z) be the space of entire function over the field of integer scalars Z. Any function $X(t) \in E(R)$ can be represented in the form of the power series

$$X(t) = \sum_{k=0}^{\infty} x_k \cdot t^k, \qquad (22)$$

where the coefficients $x_k$ satisfy the condition $\lim_{k\to\infty} \sqrt[k]{x_k} = 0$ and $t \in R$.

It is obvious that $X(n) \in E(Z)$ if $X(t) \in E(R)$. Let us define the exact difference operator $\Delta_T^k$ of the positive integer order k on the function space E(Z). The linear operator $\Delta_T^k$ will be called the exact finite difference of integer order k>0, if the following condition is satisfied: If $X(t), Y(t) \in E(R)$ and the differential equation

$$\frac{d^k Y(t)}{dt^k} = \lambda \cdot X(t) \qquad (23)$$

holds for all $t \in R$, then the difference equation

$$\Delta_T^k Y(n) = \lambda \cdot X(n) \qquad (24)$$

holds for all $n \in Z$.

In the papers (Tarasov, 2015a; Tarasov, 2016a), the exact differences of integer order have been suggested in explicit form. The exact finite difference of the first order is defined by the equation

$$\Delta_T^1 X(t) := \sum_{m=1}^{\infty} \frac{(-1)^m}{m} \cdot (X(t - Tm) - X(t + Tm)), \qquad (25)$$

where the sum implies the Cesaro or Poisson-Abel summation (Tarasov, 2016a, p. 55-56).

Equation (23) with k=1 represents the standard equation of the continuous-time accelerator. Equation (24) with the exact difference (25) represents the exact discrete analog of the standard continuous-time accelerator, which is given by equation (23) with k=1.

Exact finite difference of second and next integer orders can be defined by the recurrence formula

$$\Delta_T^{k+1} X(t) := \Delta_T^1 \left( \Delta_T^k X(t) \right). \qquad (26)$$

As a result, the exact difference of second order has the form

$$\Delta_1^2 X(t) := -\sum_{m=1}^{\infty} \frac{2 \cdot (-1)^m}{m^2} \cdot (X(t - T \cdot m) + X(t + T \cdot m)) - \frac{\pi^2}{3} \cdot X(t). \qquad (27)$$

For the arbitrary positive integer order n, the exact difference is written by the equation

$$\Delta_T^n X(t) := \sum_{m=1}^{\infty} M_n(m) \cdot (X(t - T \cdot m) + (-1)^n \cdot X(t + T \cdot m)) - M_n(0) \cdot X(t), \qquad (28)$$

where the kernel $M_n(m)$ is given by the equation

$$M_n(m) = \sum_{k=0}^{\left[\frac{n+1}{2}\right]+1} \frac{(-1)^{m+k} \cdot \Gamma(n+1) \cdot \pi^{n-2k-2}}{\Gamma(n-2k+1) \cdot m^{2k+2}} \cdot \left( (n - 2k) \cdot \cos\left(\frac{\pi n}{2}\right) + \pi \cdot m \cdot \sin\left(\frac{\pi n}{2}\right) \right) \qquad (29)$$

for $m \neq 0$, and by the expression

$$M_n(0) = \frac{\pi^n}{n+1} \cdot \cos\left(\frac{\pi n}{2}\right). \qquad (30)$$

Here we take into account that $1/\Gamma(-m) = 0$ for positive integer m.

An important characteristic property of the exact finite difference of the first order is the Leibniz rule on the space of entire functions (Tarasov, 2016a), i.e.

$$\Delta_T^1 \left( X(t) \cdot Y(t) \right) = (\Delta_T^1 X(t)) \cdot Y(t) + X(t) \cdot (\Delta_T^1 Y(t)) \qquad (31)$$

for all $X(t), Y(t) \in E(Z)$. For exact finite difference of integer order k the Leibniz rule has the form

$$\Delta_T^k (X(t) \cdot Y(t)) = \sum_{j=0}^{k} \binom{k}{j} \cdot \left( \Delta_T^{k-j} X(t) \right) \cdot \left( \Delta_T^j Y(t) \right), \qquad (32)$$

which is an exact analog of the rule for the standard derivative $d^k/dt^k$ of the integer order k.

For comparison the differences and derivatives, we give the action of the derivative and the exact difference on some elementary functions in the form of Table 2.

| f(t) | df(t)/dt | $\Delta_T^1 f(t)$ |
|---|---|---|
| $\exp(\lambda \cdot t)$ | $\lambda \cdot \exp(\lambda \cdot t)$ | $\lambda \cdot \exp(\lambda \cdot t)$ |
| $\sin(\lambda \cdot t)$ | $\lambda \cdot \cos(\lambda \cdot t)$ | $\lambda \cdot \cos(\lambda \cdot t)$ |
| $\cos(\lambda \cdot t)$ | $-\lambda \cdot \sin(\lambda \cdot t)$ | $-\lambda \cdot \sin(\lambda \cdot t)$ |
| $t^2$ | $2 \cdot t$ | $2 \cdot t$ |
| $t^3$ | $3 \cdot t^2$ | $3 \cdot t^2$ |

Table 2. Actions of derivatives and exact differences.

Note that the elementary functions, which are considered in the table, are examples of the entire functions. In the paper (Tarasov, 2016a), we prove that that the action of exact differences $\Delta_T^1$ on the space of entire function coincides with the action of first derivative. As a result, the solutions of the equations with exact differences coincide with solutions of the wide class of differential equations (Tarasov, 2016a). The equivalence of the actions of derivatives and exact differences leads to the equivalence of wide class of macroeconomic models with discrete and continuous time if the exact differences will be used. Let us demonstrate the equivalence of the continuous and discrete Harrod-Domar growth models. For this purpose we shall use the concept of an exact discrete accelerator (Tarasova and Tarasov, 2017d).

The exact difference analog of differential equation (1) of the standard accelerator has the from

$I(t) = v \cdot (\Delta_T^1 Y)(t),$  (33)

which can be written as

$I(t) = v \cdot \sum_{k=1}^{\infty} \frac{(-1)^k}{k} \cdot (Y(t - T \cdot k) - Y(t + T \cdot k)).$  (34)

Using the Newton-Leibniz theorem, equation (1) can be written in the form of the integral equation

$Y(t) = Y(0) + \frac{1}{v} \cdot \int_0^t I(\tau) d\tau.$  (35)

The exact difference analog of integral equation (35), which corresponds to (34), has the form

$Y(t) = Y(0) + \frac{1}{v} \cdot \sum_{k=1}^{\infty} \frac{Si(\pi \cdot k)}{\pi} \cdot (I(t - T \cdot k) - I(t + T \cdot k)),$  (36)

where $Si(\pi \cdot k)$ is the sine integral and $\Delta_T^1 Y(0) = 0$. In equation (36), we use the exact difference $\Delta_T^{-1}$ of the first negative order that can be considered as an exact discrete analog of the antiderivative (Tarasov, 2015a; Tarasov, 2016a), such that the relations $(\Delta_T^1 \Delta_T^{-1} X)(t) = X(t)$ and $(\Delta_T^{k+1} \Delta_T^{-1} X)(t) = (\Delta_T^k X)(t)$ are satisfied for all $X(t) \in E(Z)$.

Discrete equation, which is exact discrete analog of the Harrod-Domar model with continuous time, can be rewritten in the form

$(\Delta_T^1 Y)(t) = \lambda \cdot Y(t) - \frac{1}{v} \cdot A(t),$  (37)

where $\lambda = s/v$ and $s = 1 - c$ is the marginal propensity to save. The solution of this equation with $A(t) = A = const$ has the form

$Y(t) = A/s + (Y(0) - A/s) \cdot \exp(\lambda \cdot t).$  (38)

The fact that the function (38) is a solution of the exact-difference equation (37) can be verified by direct substitution of this function into equation (37) and using the following equalities $\Delta_T^1 \exp(\lambda \cdot t) = \lambda \cdot \exp(\lambda \cdot t)$ and $\Delta_T^1(A/s) = 0$.

Solution (38) coincides with solution (12) of equation (9) of the Harrod-Domar model with continuous time. As a result, we can state the discrete Harrod-Domar growth model, which is used exact differences, is equivalent to the continuous Harrod-Domar growth model, which is based on the differential equation.

As a result, using the Harrod-Domar growth model as an example, we proved that equations of the continuous-time models and the corresponding discrete-time models, which are based on the suggested exact differences, have the same solutions. These discrete and continuous macroeconomic models describe the same behavior of the economy.

## 6. Numerical Comparison

Let us give an illustration of the difference between the proposed approach and the standard approach by simple computer simulation of output (income) growth. We will compare the Harrod-Domar growth model with continuous time (Allen, 1960, p. 64-66), the standard Harrod-Domar growth model with discrete time (Allen, 1960, p. 74-76), and the suggested exact discretization of the Harrod-Domar model with continuous time.

The comparison of the growth in the continuous model, the exact discrete and the standard discrete models will be illustrated by simple numerical examples of the output growth, which is described by equations (12), (21), (38) with $A = 0$.

The comparison of the output growth of the continuous model (CM), the standard discrete model (SDM), and the exact discrete model (EDM) is given by Table 3. The first column specifies the growth rate of CM; the second column gives the growth rate of EDM. Note that the growth rate of CM and EDM coincides. The third column gives the growth rate of SDM. The fourth column specifies the difference between the growth rates of CM and EDM on the one hand, and the growth rates of SDM on the other hand in percentages. The fifth column describes how many times the growth in output at $t = 10 \cdot T$ is greater for the ED model in comparison with the SD model.

| CM  | EDM | SDM   | D (%) | G (times) |
|-----|-----|-------|-------|-----------|
| 0.1 | 0.1 | 0.095 | 4.69  | 1.048     |
| 0.3 | 0.3 | 0.262 | 12.54 | 1.457     |
| 0.5 | 0.5 | 0.405 | 18.90 | 2.574     |
| 0.7 | 0.7 | 0.531 | 24.19 | 5.440     |
| 0.9 | 0.9 | 0.642 | 28.68 | 13.22     |
| 1.1 | 1.1 | 0.742 | 32.55 | 35.90     |
| 1.3 | 1.3 | 0.833 | 35.93 | 106.8     |
| 1.5 | 1.5 | 0.916 | 38.91 | 342.7     |
| 1.7 | 1.7 | 0.993 | 41.57 | 1173      |
| 1.9 | 1.9 | 1.065 | 43.96 | 4242      |

Table 3. The comparison of the growth of the continuous model (CM),
the standard discrete model (SDM), and the exact discrete model (EDM).

For example, if the growth rate of EDM and CM is equal to $\lambda = 0.3$, then the growth rate of SDM is $\ln(1 + \lambda) \approx 0.262$, i.e. the growth rate of the standard discrete model is less than the growth rate of the continuous model by more than 12 percent. The growth rate of CM and EDM coincides. As a result, for example the output at $t = 10 \cdot T$ differ by almost half times in standard discrete and continuous models with $A = 0$.

If the growth rate of EDM and CM is $\lambda = 0.9$, then the growth rate of SDM is equal to $\ln(1 + \lambda) \approx 0.642$, i.e. the growth rate of the discrete model is less than the growth rate of the continuous model by more than 28 percent. In this case, for $t = 10 \cdot T$ the output growth differ by more than 13 times for standard discrete (SDM) and exact discrete models (EDM) with $A = 0$. The output growth of CM and EDM coincides.

As a result, we have that the differences of the standard discrete model from the exact discrete and continuous models can be significantly in the magnitude of output growth. Moreover the growth of the output may differ not only in several times, but also by an order of magnitude (see the fifth column (G) of Table 3).

## 7. Conclusion

A new approach to the exact discretization of the continuous-time macroeconomic models is suggested. This approach is based on the exact finite differences that are suggested in (Tarasov, 2014; Tarasov, 2015a; Tarasov, 2015b; Tarasov, 2016a; Tarasov, 2016b). These finite differences satisfy the principle of universality and the algebraic correspondence principle (Tarasov, 2015a; Tarasov, 2016a). The finite differences have a property of universality if they do not depend on the form and parameters of the considered differential equations. An algebraic correspondence means that the exact finite differences should satisfy the same algebraic relations as the derivatives. We propose the self-consistent discrete-time description of the accelerator that is based on the exact finite differences. We proved that equations of the continuous-time macroeconomic models and the corresponding discrete-time models, which are based on the suggested exact differences, can have the same solutions. These discrete and continuous economic models can describe the same behavior of the economy.

It should be noted that the proposed approach can be used for macroeconomic models with power-law memory (Tarasov and Tarasova, 2016; Tarasova and Tarasov, 2017c, 2017d; Tarasova and Tarasov, 2018a). The continuous growth models with power-law memory have been suggested in (Tarasova and Tarasov, 2016, 2017a 2017b, 2018a, 2018b; Tarasov and Tarasova, 2017), where the Caputo fractional derivatives are used. The exact fractional differences, which are suggested in (Tarasov, 2014; Tarasov, 2015b; Tarasov, 2016a; Tarasov, 2016b), allow us to propose the exact discrete-time analogs of the continuous-time equations of the accelerator and multiplier with power-law memory that are described by the Liouville fractional integrals and derivatives. The discrete macroeconomic models, which are used exact fractional differences, can be equivalent to the continuous models of processes with memory, which is described by the Liouville fractional derivatives.

We should note that additional investigations, which are based on real data, are needed to illustrate the differences between the existing methods and the suggested approach. We have proved the advantage of the proposed approach, which is based on exact finite differences, by using the well-known Harrod-Domar growth models. A comparison of the analytical solutions of the model equations and a numerical comparison of the output growth showed a significant advantage of the proposed approach to discretization of macroeconomic models with continuous time.

**References**

1. Allen, R.G.D., 1960. Mathematical Economics. Second edition. London: Macmillan. 812 p. DOI 10.1007/978-1-349-81547-0
2. Allen, R.G.D., 1967. Macro-Economic Theory. A Mathematical Treatment. London: Palgrave Macmillan. 420 p. ISBN: 978-1-349-81543-2 DOI: 10.1007/978-1-349-81541-8
3. Domar, E.D., 1946. Capital expansion, rate of growth and employment. Econometrica, 14 (2): 137-147. DOI: 10.2307/1905364
4. Domar, E.D., 1947. Expansion and employment. The American Economic Review, 37 (1): 34-55.
5. Harrod, R., 1936. An essay in dynamic theory. Economic Journal, 49 (193): 14-33. DOI: 10.2307/2225181
6. Mickens, R.E., 1988. Difference equation models of differential equations. Mathematical and Computer Modelling, 11: 528-530. DOI: 10.1016/0895-7177(88)90549-3
7. Mickens, R.E., 1993. Nonstandard Finite Difference Models of Differential Equations. Singapore: World Scientific. 264 p. ISBN: 978-981-02-1458-6
8. Mickens, R.E., 1999. Discretizations of nonlinear differential equations using explicit nonstandard methods. Journal of Computational and Applied Mathematics, 110 (1): 181-185. DOI: 10.1016/S0377-0427(99)00233-2
9. Mickens, R.E., 2000. Applications of Nonstandard Finite Difference Schemes. Singapore: World Scientific. 264 p. ISBN: 978-981-02-41
10. Mickens, R.E., 2002. Nonstandard finite difference schemes for differential equations. Journal of Difference Equations and Applications, 8 (9): 823-847. DOI: 10.1080/1023619021000000807
11. Mickens, R.E., 2005. Advances in the Applications of Nonstandard Finite Difference Schemes. Singapore: World Scientific. 664 p. ISBN: 978-981-256-404-7
12. Potts, R.B., 1982. Differential and difference equations. American Mathematical Monthly, 89 (6): 402-407.
13. Potts, R.B., 1986. Ordinary and partial difference equations. Journal of the Australian Mathematical Society B, 27 (6): 488-501. DOI: 10.1017/S0334270000005099
14. Tarasov, V.E., 2014. Toward lattice fractional vector calculus. Journal of Physics A, 47 (35): 355204. DOI: 10.1088/1751-8113/47/35/355204
15. Tarasov, V.E., 2015a. Exact discrete analogs of derivatives of integer orders: Differences as infinite series. Journal of Mathematics, 2015: 134842. DOI: 10.1155/2015/134842
16. Tarasov, V.E., 2015b. Lattice fractional calculus. Applied Mathematics and Computation, 257: 12-33. DOI: 10.1016/j.amc.2014.11.033
17. Tarasov, V.E. 2016a. Exact discretization by Fourier transforms. Communications in Nonlinear Science and Numerical Simulation, 37: 31-61. DOI: 10.1016/j.cnsns.2016.01.006

18. Tarasov, V.E., 2016b. United lattice fractional integro-differentiation. Fractional Calculus and Applied Analysis, 19 (3): 625-664. DOI: 10.1515/fca-2016-0034
19. Tarasov, V.E., 2017. Exact discretization of fractional Laplacian. Computers and Mathematics with Applications, 73 (5) 855–863. DOI: 10.1016/j.camwa.2017.01.012
20. Tarasov, V.E. and V.V. Tarasova, 2016. Long and short memory in economics: fractional-order difference and differentiation. IRA-International Journal of Management and Social Sciences, 5 (2): 327-334. DOI: 10.21013/jmss.v5.n2.p10
21. Tarasov, V.E. and V.V. Tarasova, 2017. Time-dependent fractional dynamics with memory in quantum and economic physics. Annals of Physics, 383: 579-599. DOI: 10.1016/j.aop.2017.05.017
22. Tarasova, V.V. and V.E. Tarasov, 2016. Fractional dynamics of natural growth and memory effect in economics. European Research, 12 (23): 30-37. DOI: 10.20861/2410-2873-2016-23-004
23. Tarasova, V.V. and V.E. Taraso, 2017a. Logistic map with memory from economic model. Chaos, Solitons and Fractals, 95: 84-91. DOI: 10.1016/j.chaos.2016.12.012
24. Tarasova, V.V. and V.E. Tarasov, 2017b. Economic growth model with constant pace and dynamic memory. Problems of Modern Science and Education, 2 (84): 40-45. DOI: 10.20861/2304-2338-2017-84-001
25. Tarasova, V.V. and V.E. Tarasov, 2017c. Comments on the article «Long and short memory in economics: fractional-order difference and differentiation». Problems of Modern Science and Education, 31 (113): 26-28. DOI: 10.20861/2304-2338-2017-113-002
26. Tarasova, V.V. and V.E. Tarasov, 2017d. Exact discretization of economic accelerator and multiplier with memory. Fractal and Fractional, 1 (1): Article ID: 6. DOI: 10.3390/fractalfract1010006
27. Tarasova, V.V. and V.E. Tarasov, 2018a. Concept of dynamic memory in economics. Communications in Nonlinear Science and Numerical Simulation, 55: 127-145. DOI: 10.1016/j.cnsns.2017.06.032
28. Tarasova, V.V. and V.E. Tarasov, 2018b. Dynamic intersectoral models with power-law memory. Communications in Nonlinear Science and Numerical Simulation, 54: 100-117. DOI: 10.1016/j.cnsns.2017.05.015